\documentclass[english,notitlepage,showpacs,preprintnumbers,amsmath,amssymb,aps,twocolumn,nofootinbib]{revtex4-1}
\usepackage[T1]{fontenc}
\usepackage[latin9]{inputenc}
\usepackage{array}
\usepackage{multirow}
\usepackage{amsmath}
\PassOptionsToPackage{normalem}{ulem}
\usepackage{ulem}

\makeatletter

\providecommand{\tabularnewline}{\\}

\@ifundefined{textcolor}{}
{%
 \definecolor{BLACK}{gray}{0}
 \definecolor{WHITE}{gray}{1}
 \definecolor{RED}{rgb}{1,0,0}
 \definecolor{GREEN}{rgb}{0,1,0}
 \definecolor{BLUE}{rgb}{0,0,1}
 \definecolor{CYAN}{cmyk}{1,0,0,0}
 \definecolor{MAGENTA}{cmyk}{0,1,0,0}
 \definecolor{YELLOW}{cmyk}{0,0,1,0}
}

\usepackage{amsfonts}
\usepackage{palatino}
\usepackage[]{algorithm2e}
\usepackage{MnSymbol}
\usepackage{bbm}
\usepackage{hyperref}
\usepackage{accents} 

\makeatother

\usepackage{babel}
\begin{document}

\title{Reducing multi-qubit interactions in adiabatic quantum computation.
Part 1: The ``deduc-reduc'' method and its application to quantum
factorization of numbers}

\author{Richard Tanburn$^{1,\,}$}

\email{richard.tanburn@hertford.ox.ac.uk}

\selectlanguage{english}%

\affiliation{$^{1}$Mathematical Institute, Oxford University, OX2 6GG, Oxford,
UK. }

\author{Emile Okada$^{2,\,}$}

\email{eto25@cam.ac.uk}

\selectlanguage{english}%

\affiliation{$^{2}$Department of Mathematics, Cambridge University, CB2 3AP,
Cambridge, UK. }

\author{Nikesh S. Dattani$^{3,4,\,}$}

\email{nike.dattani@gmail.com}

\selectlanguage{english}%

\affiliation{$^{3}$School of Materials Science and Engineering, Nanyang Technological
University, 639798, Singapore,}

\affiliation{$^{4}$Fukui Institute for Fundamental Chemistry, 606-8103, Kyoto,
Japan}
\begin{abstract}
Adiabatic quantum computing has recently been used to factor 56153
[Dattani \& Bryans, \href{http://arxiv.org/abs/1411.6758}{arXiv:1411.6758}]
at room temperature, which is orders of magnitude larger than any
number attempted yet using Shor's algorithm (circuit-based quantum
computation). However, this number is still vastly smaller than RSA-768
which is the largest number factored thus far on a classical computer.
We address a major issue arising in the scaling of adiabatic quantum
factorization to much larger numbers. Namely, the existence of many
4-qubit, 3-qubit and 2-qubit interactions in the Hamiltonians. We
showcase our method on various examples, one of which shows that we
can remove 94\% of the 4-qubit interactions and 83\% of the 3-qubit
interactions in the factorization of a 25-digit number with almost
no effort, \emph{without }adding any auxiliary qubits. Our method
is not limited to quantum factoring. Its importance extends to the
wider field of discrete optimization. Any CSP (constraint-satisfiability
problem), psuedo-boolean optimization problem, or QUBO (quadratic
unconstrained Boolean optimization) problem can in principle benefit
from the ``deduction-reduction'' method which we introduce in this
paper. We provide an open source code which takes in a Hamiltonian
(or a discrete discrete function which needs to be optimized), and
returns a Hamiltonian that has the same unique ground state(s), no
new auxiliary variables, and as few multi-qubit (multi-variable) terms
as possible with deduc-reduc.
\end{abstract}


\maketitle

\section{Introduction}

The quantum algorithm which has perhaps generated the most enthusiasm
about quantum computing, is Shor's algorithm for factoring integers
\cite{Shor1994,Shor1997}. However, despite being celebrated for more
than 20 years, this algorithm has still never been successfully implemented
for determining the factors of any integers without using knowledge
of the answer to the problem \cite{Smolin2013,DattaniBryans2014}.
By using knowledge of the answer to the factorization problem, one
can choose a base such that Shor's algorithm can be implemented with
fewer qubits, and by doing this, the algorithm has successfully been
implemented for factoring 15 \cite{Vandersypen2001,Lanyon2007,Lu2007,Politi2009,Martin-Lopez2012}
and 21 \cite{Lucero2012}. However, at least 8 qubits are needed for
genuinely factoring 15 with Shor's algorithm \cite{Smolin2013}, and
the largest number of qubits ever successfully used in the algorithm
was 7 \cite{Vandersypen2001}. 

Adiabatic quantum computing (AQC) has succeeded in factoring much
larger numbers with far fewer qubits, without any assumptions about
the answer to the factorization problem \cite{DattaniBryans2014}.
The largest number found so far that has been factored by the room-temperature
AQC discrete minimization experiment of \cite{Xu2012} is 56153 and
it only needed 4 qubits \cite{DattaniBryans2014}. Furthermore, quantum
annealing can be used on AQC algorithms with up to 2048 qubits, which
is enough to factor the 100-digit number RSA-100. 

Furthermore, quantum annealers have doubled in number of qubits every
year between the 4-qubit Calypso machine in 2005 and the 2048-qubit
Washington machine in 2015 (a phenomenon analogous to Moore's law
and known as ``Rose's Law''). RSA-220, which is the smallest RSA
number that has not yet been factored by any computer (whether quantum
or classical), would need only $\sim5000$ qubits to factor successfully
with the AQC algorithm of \cite{DattaniBryans2014}, which would be
well within the 8192-qubit capacity of a 2017 quantum annealer if
Rose's law continues to hold as it has done for the last 10 years. 

However, there is a major obstacle holding back such quantum annealers
with colossal numbers of qubits, from implementing the AQC factoring
algorithm. All such quantum annealing devices reported to date can
only implement AQC algorithms which have at most 2-qubit interactions
in their Hamiltonian. The algorithm which can factor RSA-220 with
$\sim5000$ qubits has a Hamiltonian with many 4-qubit and 3-qubit
interactions. In 2007 Schaller and Schutzhold devised an alternate
AQC algorithm for factoring integers, which \emph{only} has up to
2-qubit interactions in the Hamiltonian, and whose Hamiltonian's spectral
width (which is used for determining the runtime of the annealing)
increases only polynomially in the number being factored \cite{Schaller2010}. 

The problem with this AQC algorithm that uses at most 2-qubit interactions
and whose spectral width only increases polynomially, is that it requires
adding a large number of auxiliary qubits (27\,225 for RSA-100, and
133\,225 for RSA-220). In this paper we present a method, called
``deduc-reduc'' for reducing the number of 4-qubit and 3-qubit terms
in the AQC Hamiltonian \emph{without} adding auxiliary qubits. 

While the motivation for this work was the problem of factoring numbers
using AQC, deduc-reduc can be used for a wider range of discrete optimization
problems such as constraint satisfaction problems (CSPs) and pseudo-Boolean
optimization problems, and problems arising in the field of artificial
intelligence/neural networks \cite{Altaisky2015}. All of these problems
can benefit from quantum annealing because of the capability for quadratic
speed-up over simulated annealing \cite{Somma2008}. Quantum annealing
has also recently been shown to be able to provide exponential speed-up
for some specific discrete optimization problems \cite{Boixo2015},
and the potential for speeding up more general discrete optimization
problems is an active area of research. 

It has also been shown that AQC can simulate any circuit-based quantum
algorithm with only polynomial overhead \cite{Mizel2007,Aharonov2008}.
Therefore, apart from the exponential speed-up offered for many classes
of problems, AQC can provide polynomial speed-up over the best classical
3-SAT algorithms \cite{Ambainis2005} and other CSPs \cite{Cerf2000},
and can provide exponential speed-up for combinatorial optimization
problems for which an approximate solution is sufficient \cite{Farhi2014}.

\section{A quick example}

It has been shown with concrete examples in many papers \cite{DattaniBryans2014,Xu2012,Schaller2010,Burges2002}
that integer factorization is equivalent to solving a system of binary
equations such as:

\vspace{-6.5mm}

\begin{eqnarray}
x_{1}+x_{2}+x_{3} & = & 1\label{eq:example1SystemEq1-1}\\
x_{1}x_{4}+x_{2}x_{5} & = & x_{3}\label{eq:example1SystemEq2-1}\\
x_{1}+2x_{2} & = & x_{3}+2x_{4}\label{eq:example1SystemEq3-1}
\end{eqnarray}
which {\small{}}is equivalent to finding the minimum of an objective
function formed by subtracting the right hand side of each equation
from its left, squaring, and summing (and remembering that the square
of a binary variable is itself): \vspace{-6.5mm}

{\small{}
\begin{eqnarray}
H & = & (x_{1}\negthinspace+\negthinspace x_{2}\negthinspace+\negthinspace x_{3}\negthinspace-\negthinspace1)^{2}\negthinspace+\negthinspace(x_{1}x_{4}\negthinspace+\negthinspace x_{2}x_{5}\negthinspace-\negthinspace x_{3})^{2}\negthinspace+\\
 &  & (x_{1}\negthinspace+\negthinspace2x_{2}\negthinspace-\negthinspace x_{3}\negthinspace-\negthinspace2x_{4})^{2}\\
 & = & 2x_{1}x_{2}x_{4}x_{5}\negthinspace-\negthinspace2x_{1}x_{3}x_{4}\negthinspace-\negthinspace2x_{2}x_{3}x_{5}\negthinspace-\negthinspace2x_{2}x_{3}\negthinspace+\negthinspace6x_{1}x_{2}\label{eq:example1objFunc-1}\\
 &  & -3x_{1}x_{4}\negthinspace-\negthinspace8x_{2}x_{4}\negthinspace+\negthinspace x_{2}x_{5}\negthinspace+\negthinspace3x_{2}\negthinspace+\negthinspace4x_{3}x_{4}\negthinspace+\negthinspace x_{3}\negthinspace+\negthinspace4x_{4}\negthinspace+\negthinspace1
\end{eqnarray}
}{\small \par}

If we represent each variable $x_{i}$ as a qubit based on a Pauli
spin matrix in its own Hilbert space:

\begin{equation}
x_{i}=\frac{1}{2}\left(1-\sigma_{z}^{(i)}\right)^{2}=\begin{pmatrix}0 & 0\\
0 & 1
\end{pmatrix},
\end{equation}
and each term of Eq. \ref{eq:example1objFunc-1} using left Kronecker
products (or ``tensor products'') in the Hilbert space of all qubits,
for example:

\begin{widetext}

\begin{eqnarray}
2x_{1}x_{2}x_{4}x_{5} & = & 2x_{1}x_{2}\openone_{3}x_{4}x_{5}=2\begin{pmatrix}0 & 0\\
0 & 1
\end{pmatrix}\otimes\begin{pmatrix}0 & 0\\
0 & 1
\end{pmatrix}\otimes\begin{pmatrix}1 & 0\\
0 & 1
\end{pmatrix}\otimes\begin{pmatrix}0 & 0\\
0 & 1
\end{pmatrix}\otimes\begin{pmatrix}0 & 0\\
0 & 1
\end{pmatrix},\\
 & = & {\rm diag}(0,0,0,0,0,0,0,0,0,0,0,0,0,0,0,0,0,0,0,0,0,0,0,0,0,0,0,2,0,0,0,2),
\end{eqnarray}
we see quickly that the objective function of Eq. \ref{eq:example1objFunc-1}
can be represented as a diagonal 32$\times$32 Hamiltonian matrix:

\[
H={\rm diag(1,1,5,5,2,2,10,10,4,5,0,1,3,2,3,2,1,1,2,2,2,2,5,5,10,11,3,6,9,8,4,5)}.
\]

The lowest eigenvalue is clearly 0, and it is at position:
\begin{equation}
{\rm diag(0,0,0,0,0,0,0,0,0,0,1,0,0,0,0,0,0,0,0,0,0,0,0,0,0,0,0,0,0,0,0,0)}=|x_{1}=0,x_{2}=1,x_{3}=0,x_{4}=1,x_{5}=0\rangle.
\end{equation}

\end{widetext}

This may seem like an unnecessary extra complication to find the minimum,
especially since if we have $N$ variables, it involves representing
all 2$^{N}$ values of $H$ and then searching for the lowest one.
However, since the lowest eigenvalue of a Hamiltonian corresponds
to its ground state, the minimum of a $2^{N}$ variable system can
be found simply by connecting $N$ qubits together with appropriate
energies and coupling strengths, then finding the ground state of
this system. This avoids having to evaluate the function $2^{N}$
times. But it is \emph{much }harder to couple three qubits together
as in the cubic term $2x_{1}x_{2}x_{3}$, than it is to couple two
qubits together as in the quadratic terms, which is of course still
harder than not having to enforce any couplings (but if only linear
terms exist, the problem can be solved trivially by setting all negative
terms to 1 and positive terms to 0). Likewise, even if solving this
problem on a classical computer, multi-variable (or multi-qubit) terms
make the problem \emph{much} more difficult, and it is desirable to
eliminate any terms containing more than 2 variables, since this would
transform the problem into a quadratic unconstrained Boolean optimization
(QUBO) problem, whereby we can use many beautiful algorithms and results
such as the fact that the solution can be found in polynomial time
if all quadratic terms are negative \cite{Nemhauser1981,Orlin2007,Jegelka2011,Boros2002}.
Furthermore, it is also desirable to eliminate as many quadratic terms
as possible.

\subsection{The method}

We now demonstrate how deduc-reduc can be used to eliminate multi-qubit
interactions (or high-order terms). 

From Eq. \ref{eq:example1SystemEq1-1} alone we can make the \emph{deduction}s:
\begin{eqnarray}
x_{1}x_{2} & \negthinspace=\negthinspace x_{2}x_{3}\negthinspace=\negthinspace x_{3}x_{1}\negthinspace=\negthinspace & 0\label{eq:example1Const1-1}
\end{eqnarray}
since the LHS of Eq. \ref{eq:example1SystemEq1-1} cannot be greater
than 1, and this would not be true if any of the products in Eq. \ref{eq:example1Const1-1}
were 1.

\subsubsection*{Naive substitution}

Suppose that we substituted Eq. \ref{eq:example1Const1-1} back into
$H$:{\small{}
\begin{eqnarray}
H & = & -3x_{1}x_{4}\negthinspace-\negthinspace8x_{2}x_{4}\negthinspace+\negthinspace x_{2}x_{5}\negthinspace+\negthinspace3x_{2}\negthinspace+\negthinspace4x_{3}x_{4}\negthinspace+\negthinspace x_{3}\negthinspace+\negthinspace4x_{4}\negthinspace+\negthinspace1.
\end{eqnarray}
}Then we find that while $H(0,1,0,1,0)=0$ matches the original Hamiltonian
as required, we have also introduced a new state with $H=-3$ and
another for which $H=-2$.  This is a problem, because if the minimum
is not 0, then the ground state does \emph{not} encode the solution
to Eqs \ref{eq:example1SystemEq1-1}-\ref{eq:example1SystemEq3-1},
which is the integer factorization problem we want to solve.

\subsubsection*{Deduc-reduc}

Instead, we can look at each term we would like to substitute Eq.
\ref{eq:example1Const1-1} into, and calculate an associated error
term.

We want an error term such that it is equal to $0$ if $x_{1}x_{2}=0$
and strictly greater than $0$ if $x_{1}x_{2}\neq0$. In this particular
example, we can take $x_{1}x_{2}$ to be our error term and similarly
for $x_{2}x_{3}$ and $x_{3}x_{1}$. We will now work through the
rest of this example in order to illustrate the method, but a general
formulation can be found in Section \ref{sec:General-Treatment}. 

We have three cubic/quartic terms to consider:
\begin{enumerate}
\item $2x_{1}x_{2}x_{4}x_{5}$. Here we make use of the judgment $x_{1}x_{2}=0$.
For any state, $2x_{1}x_{2}x_{4}x_{5}$ can only evaluate to $0$
or $2$. In the former case, a straight substitution does not make
any difference to the value of $H$ and setting the term to $0$ is
permissible. If the term evaluates to $2$ (i.e. all variables are
equal to $1$) and we perform a straight substitution, we are taking
$2$ away from the energy of the state. This may introduce new states
with $H=0$ or even negative valued states. To compensate for this,
we add $2x_{1}x_{2}$ back on to the original Hamiltonian to preserve
the original value of $H$. We have then made the transformation $2x_{1}x_{2}x_{4}x_{5}\rightarrow2x_{1}x_{2}$.
While this step may add 2 to the original energy of a state, the important
detail is that we know this will not happen for any ground states.
\item $-2x_{1}x_{3}x_{4}$. This can only take the values $0$ or $-2$.
In the first case, a straight substitution makes no difference. In
the second, a straight substitution for $0$ would add $2$ to the
value of the incorrect state, which is safe to do, (it is even desirable,
since increasing the energy gap between the ground and excited states
can allow the ground state to be found quicker by annealing or adiabatic
evolution methods, and the power to manipulate the energy landscape
like this is the subject of our forthcoming paper \cite{Lunt2015}).
Hence we can make the transformation $-2x_{1}x_{3}x_{4}\rightarrow0$.
\item $-2x_{2}x_{3}x_{5}$. As with case 2, we make the transformation $-2x_{2}x_{3}x_{5}\rightarrow0$.
\end{enumerate}
Bringing the above substitutions together we get a Hamiltonian with
the \emph{same} number of qubits as before, but has \emph{no }4-qubit
(quartic) nor 3-qubit (cubic) terms, and whose only ground state is
precisely the ground state of the original Hamiltonian:

{\small{}}{\small \par}

{\small{}
\begin{eqnarray}
H & = & 8x_{1}x_{2}\negthinspace-\negthinspace2x_{2}x_{3}\negthinspace-\negthinspace3x_{1}x_{4}\negthinspace-\negthinspace8x_{2}x_{4}\negthinspace+\negthinspace x_{2}x_{5}\negthinspace+\label{eq:toyExampleReduced}\\
 &  & 3x_{2}\negthinspace+\negthinspace4x_{3}x_{4}\negthinspace+\negthinspace x_{3}\negthinspace+\negthinspace4x_{4}\negthinspace+\negthinspace1.
\end{eqnarray}
}{\small \par}

A quick numerical check shows that our new Hamiltonian has a single
zero at exactly the original solution as required. We have therefore
used our \emph{\uline{deductions}}\emph{ }in Eq. \ref{eq:example1Const1-1}
to do a \emph{\uline{reduction}} of the 4-qubit and 3-qubit terms.

\subsubsection*{}

\subsubsection*{Removing quadratic terms}

Notice that when the coefficient of a term was negative, we were able
to perform a straight substitution, as doing so only added to the
energy of a state and no error term was needed. This can also be used
to eliminate quadratic terms like $-2x_{2}x_{3}$, to provide extra
simplification. Our final Hamiltonian is then:

{\small{}
\begin{eqnarray}
\negmedspace\negmedspace\negmedspace\negmedspace\negmedspace H & \negmedspace=\negmedspace & 8x_{1}x_{2}\negthinspace-\negthinspace3x_{1}x_{4}\negthinspace-\negthinspace8x_{2}x_{4}\negthinspace+\negthinspace x_{2}x_{5}\negthinspace+\negthinspace3x_{2}\negthinspace+\negthinspace4x_{3}x_{4}\negthinspace+\negthinspace x_{3}\negthinspace+\negthinspace4x_{4}\negthinspace+\negthinspace1.
\end{eqnarray}
}{\small \par}

{\small{}Note that in general we can remove quadratic terms by following
the method described in Section \ref{sec:General-Treatment}. However,
this is difficult in practice because any deduction must be at least
quadratic (else we perform a straight substitution) and requires an
appropriate error term.}{\small \par}

\subsection{Performance of deduc-reduc on large examples}

We now consider the performance of deduc-reduc on Hamiltonians whose
ground states encode the prime factors to larger numbers, displayed
in Table \ref{tab:the3examples}.

\begin{table}
\protect\caption{The three examples used to illustrate the power of deduc-reduc for
removing multi-qubit interactions.\label{tab:the3examples}}

\begin{centering}

\par\end{centering}

\begin{centering}
\begin{tabular*}{1\columnwidth}{@{\extracolsep{\fill}}ccc}
\hline 
\noalign{\vskip2mm}
\multirow{2}{*}{Example} & \multirow{2}{*}{Product} & Length of product \tabularnewline
 &  & in binary\tabularnewline[2mm]
\hline 
\noalign{\vskip2mm}
1 & $455937533473$ & 40\tabularnewline
2 & $292951160076082381$ & 60\tabularnewline
3 & $1208925727750433490141601$ & 80\tabularnewline[2mm]
\hline 
\end{tabular*}
\par\end{centering}

\centering{}\rule[0.5ex]{1\columnwidth}{0.5pt}
\end{table}

For each example, we generate the carry equations that express the
factorization and then form the Hamiltonian as in \cite{Burges2002,DattaniBryans2014}:
this is our starting Hamiltonian $H_{0}$. Next, we perform simple
logical deductions to solve for the most obvious variables (as in
Eq. \ref{eq:example1Const1-1} and the Supplementary Material of \cite{Xu2012}),
using no enumeration of solutions or deduc-reduc machinery. We call
this reduced Hamiltonian $H_{1}$.

Finally we perform searches through the state space by enumerating
plausible solutions and looking for patterns that can be used by deduc-reduc.
We perform a breadth-first search that halts when it has found more
than $n$ plausible solutions. We call the Hamiltonian formed using
deduc-reduc with the deductions from such a search $H_{n}$. 

 Enumerating all possible solutions is akin to doing a brute force
search for the minimum, so of course it would make no sense to run
this algorithm for $n$ anywhere close to the total size of the search
space $2^{N}$ where $N$ is the number of qubits in the Hamiltonian.
However, in the examples from Table \ref{tab:the3examples}, we show
in Table \ref{tab:bigExample1} that with $n$ vastly smaller than
$2^{N}$, we can still eliminate a significant number of multi-qubit
interactions %
\footnote{While the factorization problem can be directly solved by brute force
using $2^{m}<2^{N}$ trials, where $m$ is the bit-length of the number
we wish to factorize, most discrete optimization problems will not
have this feature, so the brute force search to solve the problem
will generally require $2^{N}$ trials where $N$ is the number of
variables. Furthermore, the number of states in our search is still
vastly fewer than even $2^{m}$.%
}. No deduc-reduc example we present takes more than a few hours of
CPU time on an average laptop, and 1000 states takes just a few seconds.
The code is written in Python with scope for optimization and parallelization.

\begin{table*}
\protect\caption{Performance of deduc-reduc on the three examples from Table \ref{tab:the3examples}.\label{tab:bigExample1}}

\begin{centering}
\begin{tabular*}{1\textwidth}{@{\extracolsep{\fill}}ccccccc}
\hline 
\noalign{\vskip2mm}
\multicolumn{7}{c}{$455937533473=524309\times869597=10000000000000010101\times11010100010011011101$}\tabularnewline[2mm]
\hline 
\noalign{\vskip2mm}
Hamiltonian & \# of qubits & \# of deductions & 4-qubit terms & 3-qubit terms & 2-qubit terms & 1-qubit terms\tabularnewline[2mm]
\hline 
\noalign{\vskip2mm}
Original, $H_{0}$ & 174 & - & 1785 & 3318 & 1783 & 150\tabularnewline
Simple Judgments, $H_{1}$ & 148 & - & 1750 & 2915 & 1407 & 128\tabularnewline
Reduction with 100 states, $H_{100}$ & 146 & 88 & 1645 & 2828 & 1404 & 128\tabularnewline
Reduction with 1000 states, $H_{1000}$ & 144 & 121 & 1645 & 2794 & 1411 & 126\tabularnewline
Reduction with 10\,000 states, $H_{10\,000}$ & 141 & 131 & 1645 & 2704 & 1362 & 123\tabularnewline
Reduction with 100\,000 states, $H_{100\,000}$ & 138 & 242 & 1645 & 2591 & 1370 & 120\tabularnewline[2mm]
\hline 
\hline 
\noalign{\vskip2mm}
\multicolumn{7}{c}{$292951160076082381=539152967\times543354443=100000001000101101001001000111\times100000011000101110111001001011$}\tabularnewline[2mm]
\hline 
\noalign{\vskip2mm}
Hamiltonian & \# of qubits & \# of deductions & 4-qubit terms & 3-qubit terms & 2-qubit terms & 1-qubit terms\tabularnewline[2mm]
\hline 
\noalign{\vskip2mm}
Original, $H_{0}$ & 294 & - & 6930 & 8816 & 3541 & 268\tabularnewline
Simple Judgments, $H_{1}$ & 200 & - & 3686 & 5061 & 2222 & 180\tabularnewline
Reduction with 100 states, $H_{100}$ & 199 & 100 & 3388 & 4942 & 2193 & 180\tabularnewline
Reduction with 1000 states, $H_{1000}$ & 199 & 160 & 3388 & 4924 & 2192 & 180\tabularnewline
Reduction with 10\,000 states, $H_{10\,000}$ & 189 & 111 & 3388 & 4860 & 2074 & 173\tabularnewline
Reduction with 100\,000 states, $H_{100\,000}$ & 185 & 129 & 3388 & 4747 & 1992 & 169\tabularnewline[2mm]
\hline 
\hline 
\noalign{\vskip2mm}
\multicolumn{7}{c}{$1208925727750433490141601=1099511555521\times1099511616481$}\tabularnewline
\multicolumn{7}{c}{$1111111111111111111111101110010111000001\times1111111111111111111111111101001111100001$}\tabularnewline[2mm]
\hline 
\noalign{\vskip2mm}
Hamiltonian & \# of qubits & \# of deductions & 4-qubit terms & 3-qubit terms & 2-qubit terms & 1-qubit terms\tabularnewline[2mm]
\hline 
\noalign{\vskip2mm}
Original, $H_{0}$ & 430 & - & 17575 & 17762 & 5891 & 382\tabularnewline
Simple Judgments, $H_{1}$ & 367 & - & 16133 & 16005 & 5256 & 367\tabularnewline
Reduction with 100 states, $H_{100}$ & 333 & 56 & 8469 & 14203 & 6881 & 327\tabularnewline
Reduction with 1000 states, $H_{1000}$ & 257 & 289 & 1449 & 3649 & 6180 & 257\tabularnewline
Reduction with 10\,000 states, $H_{10\,000}$ & 253 & 285 & 1449 & 3565 & 6106 & 253\tabularnewline
Reduction with 100\,000 states, $H_{100\,000}$ & 236 & 236 & 1142 & 2961 & 5732 & 236\tabularnewline[2mm]
\hline 
\end{tabular*}
\par\end{centering}

\centering{}\rule[-0.5ex]{1\textwidth}{0.5pt}
\end{table*}

In Example 1 we are unable to make a big difference in the number
of multi-qubit interactions because both the simple judgments and
local searches are unable to find many solutions or deductions that
are useful for deduc-reduc. In Example 2, the large improvement in
the qubit profile comes from the power of simple judgments, which
leaves little `easy' work for the search and deduc-reduc to do afterwards.
The remarkable success of deduc-reduc in Example 3 stems from the
effectiveness of the breadth-first search's ability to find patterns.
The factors that affect the performance of the simple judgments and
the breadth-first search are a potential area for future research
though we believe it is related to the distribution of $0$s and $1$s
in the binary representations of the factors.

\section{General treatment\label{sec:General-Treatment}}

\subsection{General formulation}

Let $\underline{x}$ be a binary string of length $n$. Suppose we
have a general Hamiltonian $H(\underline{x})$ and a deduction $f(\underline{x})=g(\underline{x})$
that is true for all ground states, and $f,g$ are polynomials with
the degree of $g$ strictly less than the degree of $f$. Suppose
futher that we are able to write $H(\underline{x})=q(\underline{x})f(\underline{x})+r(\underline{x})$
for some polynomials $q\neq0$ and $r$. Then substitution of $f$
for $g$ would give a new polynomial with fewer high order terms.

To allow us to perform this substitution and preserve the global minima,
we add an error term $\lambda C(\underline{x})$ of sufficient magnitude
to compensate for the substitution. If we start with the deduction
``$f(\underline{x})=g(\underline{x})$ for all minima $\underline{x}$''
and form a corresponding error term $C=(f-g)^{2}$, we define the
new polynomial: 
\begin{eqnarray}
H'(\underline{x}) & = & q(\underline{x})g(\underline{x})\negthinspace+\negthinspace r(\underline{x})\negthinspace+\negthinspace\lambda C(\underline{x})\label{eq:deducReducForm}\\
 & = & q(\underline{x})g(\underline{x})\negthinspace+\negthinspace r(\underline{x})\negthinspace+\negthinspace\lambda(f\negthinspace-\negthinspace g)^{2}
\end{eqnarray}
where $|q(\underline{x})|\leq\lambda$ for all states $\underline{x}$.
We claim that $H'$ has exactly the same minima as $H$, and no new
minima, and we prove it in the next sub-section.

\subsection{Proof of formulation}

To show that this $H^{\prime}(x)$ has the required properties, we
condition on whether or not a state $\underline{x}$ is a ground state
of the original Hamiltonian.
\begin{enumerate}
\item If $\underline{x}$ is a ground state then $f(\underline{x})\negmedspace=\negmedspace g(\underline{x})$
and $C(\underline{x})\negmedspace=\negmedspace(f\negmedspace-\negmedspace g)^{2}\negmedspace=\negmedspace0$
so we have 
\begin{eqnarray}
H'(\underline{x}) & = & q(\underline{x})g(\underline{x})\negthinspace+\negthinspace r(\underline{x})\negthinspace+\negthinspace\lambda C(\underline{x})\\
 & = & q(\underline{x})f(\underline{x})\negthinspace+\negthinspace r(\underline{x})\\
 & = & H(\underline{x})
\end{eqnarray}
 as required.
\item If $\underline{x}$ is not a solution, we have that $|q(\underline{x})|\leq\lambda$,
by construction, and $|g(\underline{x})-f(\underline{x})|\leq(g(\underline{x})-f(\underline{x}))^{2}$,
as we are dealing with polynomials taking integer values. Hence 
\begin{equation}
0\leq q(\underline{x})(g(\underline{x})\negthinspace-\negthinspace f(\underline{x}))\negthinspace+\negthinspace\lambda(g(\underline{x})\negthinspace-\negthinspace f(\underline{x}))^{2}.
\end{equation}
\vspace{2mm}
Then, 
\begin{eqnarray}
H'(\underline{x}) & = & q(\underline{x})g(\underline{x})\negthinspace+\negthinspace r(\underline{x})\negthinspace+\negthinspace\lambda C(\underline{x})\\
 & = & (q(\underline{x})f(\underline{x})\negthinspace+\negthinspace r(\underline{x}))\negthinspace+\negthinspace q(\underline{x})(g(\underline{x})\negthinspace-\negthinspace f(\underline{x}))\negthinspace+\\
 &  & \lambda C(\underline{x})\\
 & = & \negmedspace H(\underline{x})\negmedspace+\negmedspace q(\underline{x})(g(\underline{x})\negmedspace-\negmedspace f(\underline{x}))\negmedspace+\negmedspace\lambda(g(\underline{x})\negmedspace-\negmedspace f(\underline{x}))^{2}\\
 & > & 0\nonumber 
\end{eqnarray}
since $H(\underline{x})>0$.
\end{enumerate}
Hence $H(\underline{x})=0$ if and only if $H'(\underline{x})=0$.

\subsection{Calculation of $\lambda$}

While maximizing an arbitrary integer valued polynomial in $n$ variables
might seem suspiciously similar to the original problem, we only need
an upper bound. Such a bound can quickly be found by the triangle
inequality. However, this might lead to much larger coefficients than
are needed. It is perfectly permissible to split up $q(\underline{x})$
and apply repeatedly, as we did in our example to find Hamiltonian
\ref{eq:toyExampleReduced}.

In the most general case, $\lambda$ can be picked even more precisely
depending on the $f,g,C$ and $q$, as long as the following inequality
is satisfied: 
\begin{equation}
0\leq q(\underline{x})(g(\underline{x})\negthinspace-\negthinspace f(\underline{x}))+\lambda C(\underline{x})\,\mbox{for all states }\underline{x}.\label{eq:lambdaConst}
\end{equation}

\section{Conclusion}

For any discrete optimization problem, we have presented a method
(called ``deduc-reduc'') that can re-formulate the objective function
such that without adding auxiliary variables, the new objective function
has fewer high-order terms. Any input that minimizes the original
objective function also minimizes the new one, and no new inputs can
minimize the objective function. This can be used to reduce a high-order
objective function into an easier one, such as a quadratic Boolean
optimization problem (QUBO). On our examples for integer factorization,
we were able to remove thousands of quartic and cubic terms in a few
seconds on a laptop. We provide an open source code with sample input
files in \cite{Tanburn2015}. 

The method we presented can also be used to turn a target Hamiltonian
in adiabatic quantum computation (AQC), without adding auxiliary qubits,
into one with equivalent unique ground states, but with only 2-qubit
terms (for implementations using, for example SQUID-based quantum
annealing devices that cannot yet couple more than 2 qubits \cite{Ronnow2014})
or low enough order terms for other AQC-like implementations such
as NMR-based devices which can implement 3- and 4-qubit interactions
with some more effort \cite{Xu2012}.

The method we presented here can also be used to manipulate the energy
landscape of a Hamiltonian without changing the unique ground states.
Particularly, it can be used to move low-lying excited states (local
minima of the objective function) away from the ground state (global
minimum of the objective function) and to adjust the Hamiltonian's
spectral width (range of the objective function) without adding more
qubits. This would allow for adjusting the size of the spectral gap
between the ground and first excited states of the target Hamiltonian,
and therefore it can be used to decrease the runtime of the adiabatic
quantum computation, and to also speed-up conventional discrete optimization
calculations such as simulated annealing. This is the subject of a
forthcoming publication \cite{Lunt2015}.

Finally, this is the first paper of a 2-part series on techniques
for reducing multi-variable terms in discrete optimization problems.
The second method is called ``split-reduc'' and is presented in
\cite{Okada2015} with examples focused on the quadratization of Hamiltonians
used for AQC-based determination of Ramsey numbers \cite{Bian2013,Gaitan2012}.

\section*{Acknowledgements}

We gratefully thank Oliver Lunt of Oxford University's Trinity College
for careful proofreading of the manuscript.

\end{document}